\documentclass[mathleft]{an}
\usepackage{graphicx}
\usepackage{times}
\usepackage[T1]{fontenc}

\begin{document}

\Pagespan{1}{}
\Yearpublication{}%
\Yearsubmission{}%
\Month{}%
\Volume{}%
\Issue{}%

\title{The Monoceros radio loop: temperature, brightness, spectral index and distance}

\author{V. Borka Jovanovi\'{c}\inst{1}\thanks{Corresponding author:
  \email{vborka@vinca.rs}\newline}
\and D. Uro\v{s}evi\'{c} \inst{2}\thanks{
\email{dejanu@matf.bg.ac.yu}}}

\titlerunning{The Monoceros radio loop}
\authorrunning{V. Borka Jovanovi\'{c} and D. Uro\v{s}evi\'{c}}

\institute{Laboratory of Physics (010), Vin\v{c}a Institute of
Nuclear Sciences, P.O. Box 522, 11001 Belgrade, Serbia \and
Department of Astronomy, Faculty of Mathematics, University of
Belgrade, Studentski trg 16, 11000 Belgrade, Serbia}

\received{2009 January 12} \accepted{2009 April 14} \publonline{}


\abstract{In this paper we estimated the temperatures and
brightnesses of the Monoceros radio loop at 1420, 820 and 408 MHz.
Linear spectrum is estimated for mean temperatures versus frequency
between 1420, 820 and 408 MHz. The spectral index of Monoceros loop
is also obtained. The brightness temperatures and surface
brightnesses of the loop are computed using data taken from
radio-continuum surveys at the three frequencies. The spectral index
of the loop is also obtained from $T-T$ plots between 1420 - 820,
1420 - 408 and 820 - 408 MHz. The obtained results confirm
non-thermal origin of the Monoceros radio loop.}

\maketitle

\section{Introduction}

The Monoceros filamentary loop nebula was suggested to be a
supernova remnant (SNR) by \cite{davi63} on the basis of 237 MHz
observations. This shell source of radio emission was suspected
already by \cite{davi63} to be a SNR, while \cite{gebe72} developed
this idea into a more detailed study. It was considered as an object
similar to major loops when Spoelstra 1973 included it in his study
of galactic loops as supernova remnants expanding in the local
galactic magnetic field.

\cite{grah82} found that the constellation of Mo\-noceros is
remarkably rich in extended Galactic radio sour\-ces. Large parts of
it have been mapped in the radio continuum over a wide range of
frequencies. Monoceros Nebula can be found in a catalog of Galactic
SNRs listed as G205.5+0.5 (\cite{gree04,gree06}). It is S-type
remnant (shell remnant), characterized by diffuse, shell-like
emission with steep radio spectrum and size of $\sim$220$^\prime$.
Shell type supernova remnants are believed to be particle
accelerators to energy up to a few hundred TeV, and it is shown that
Monoceros loop is a good candidate for acceleration of particles
(\cite{fias07}).

The aim of this paper is to calculate the average brightness
temperatures and surface brightnesses of the Monoceros radio loop at
1420, 820 and 408 MHz and to study how these results compare with
previous result (\cite{uros98}; \cite{jmt96}) and with current
theories of supernova remnant evolution. These theories predict that
SNRs are at radio frequencies predominantly non-thermal sources
which are sp\-reading inside of the hot and low density bubbles made
by former supernova explosions or by strong stellar winds (see
\cite{salt83}; \cite{mcke77} and references there\-in).

Spectrum (temperature versus frequency) have been pl\-otted and this
result is used to determine spectral index of the Monoceros loop.

\section{Analysis}

\subsection{Data}

In this paper we used observations from several radio-conti\-nuum
surveys given in "Flexible Image Transport System" ($FITS$) format
which are available on MPIfR's Survey Sa\-mpler
("Max-Planck-Institut f\"{u}r Radioastronomie", Bonn). This is an
online service (http://www.mpifr-bonn.mpg.de/ survey.html), which
allows users to pick a region of the sky and obtain images and
observed data (in $FITS$ format) at a number of wavelengths. User
can choose a coordinate system, projection type and a survey. The
radio continuum surveys at 1420, 820 and 408 MHz provided the
database for computing brightness temperatures ($T_\mathrm{b}$). The
1420-MHz Stockert survey (\cite{reic86}) has the angular resolution
of 35$^\prime$, the 820-MHz Dwingeloo survey (\cite{berk72})
1$^\circ$.2 and the 408-MHz all-sky survey (\cite{hasl82})
0$^\circ$.85. The corresponding observations are given at the
following rates for both $l$ and $b$: $\frac{1^\circ}{4}$ at 1420
MHz, $\frac{1^\circ}{2}$ at 820 MHz and $\frac{1^\circ}{3}$ at 408
MHz. The effective sensitivities are about 50 mK T$_b$, 0.20 K and
about 1.0 K, respectively.

\subsection{Method}

We extracted observed brightness temperatures from $FITS$ format
into ASCII data files and after that, these data files have been
processed by our software, i.e. we have developed several programs
in C and FORTRAN in order to obtain results presented in this paper.

\begin{figure*}
\centering
\includegraphics[width=0.38\textwidth]{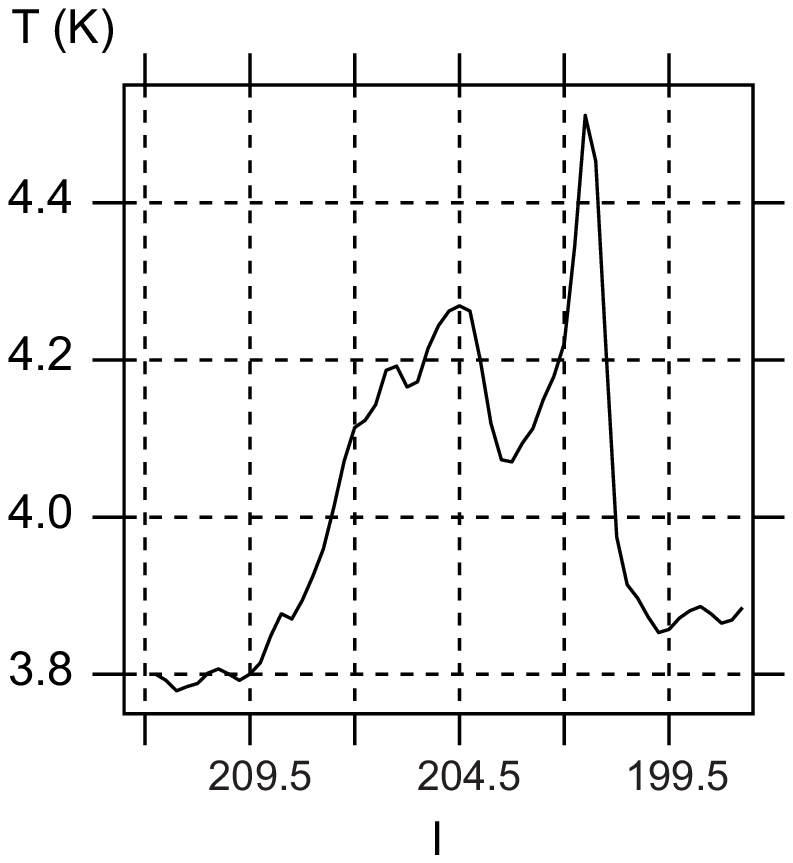}
\includegraphics[width=0.38\textwidth]{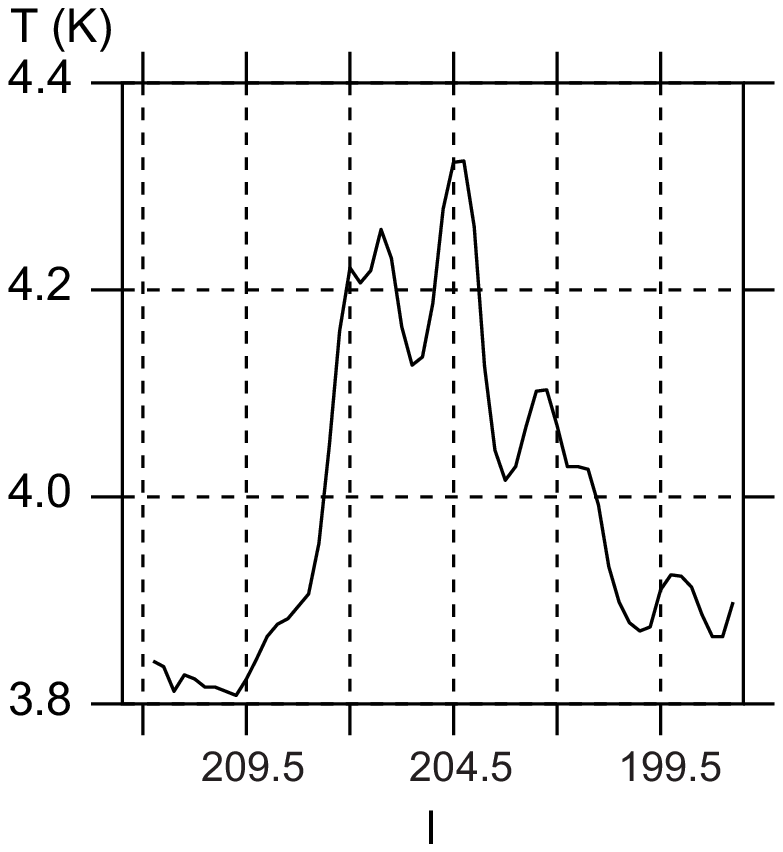}
\includegraphics[width=0.38\textwidth]{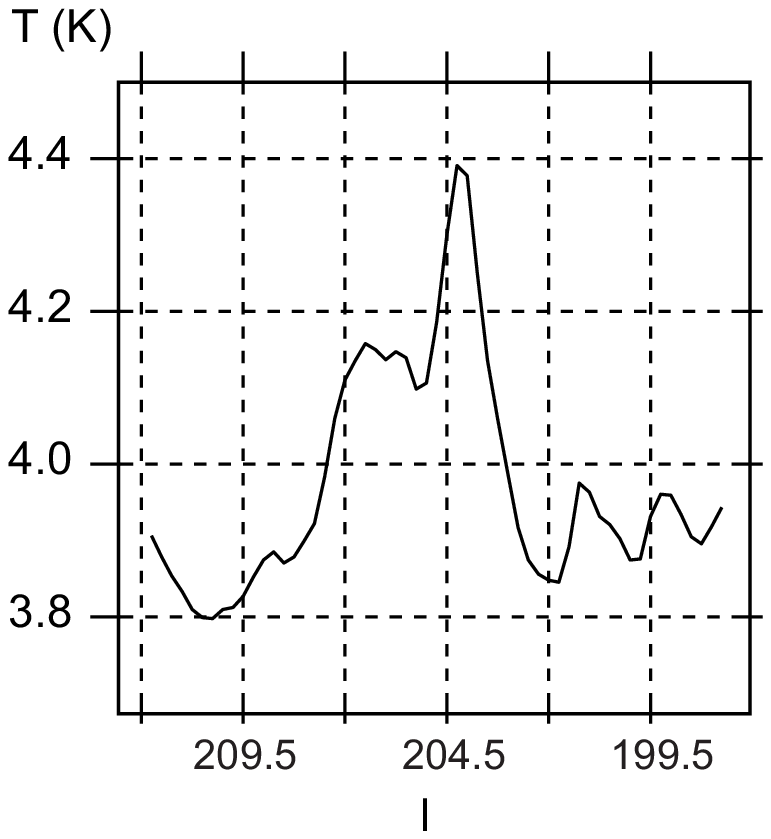} \\
\includegraphics[width=0.38\textwidth]{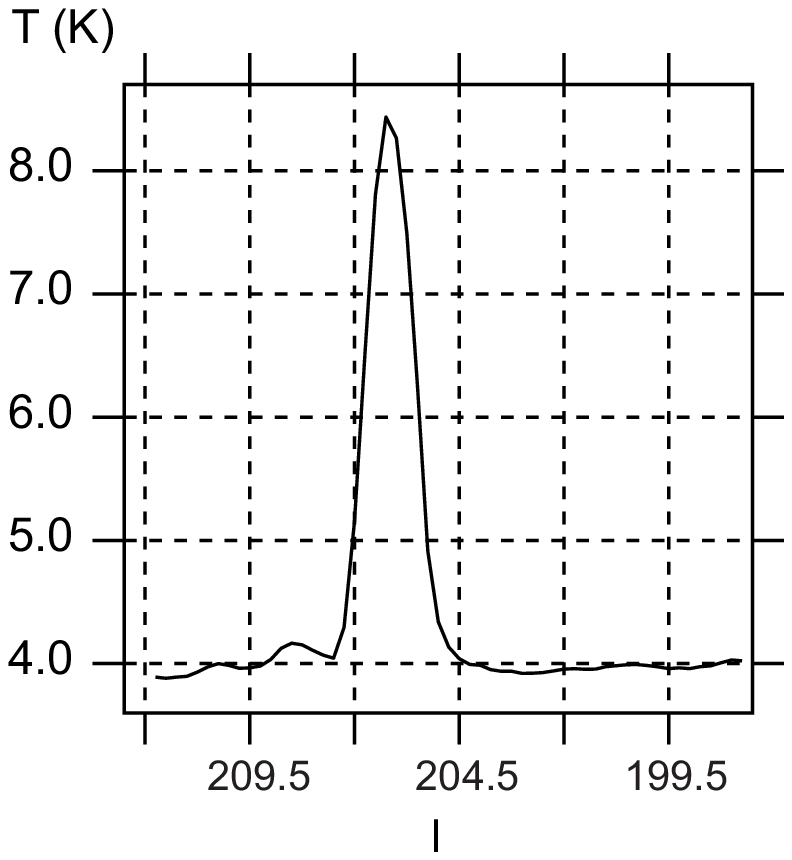}
\includegraphics[width=0.38\textwidth]{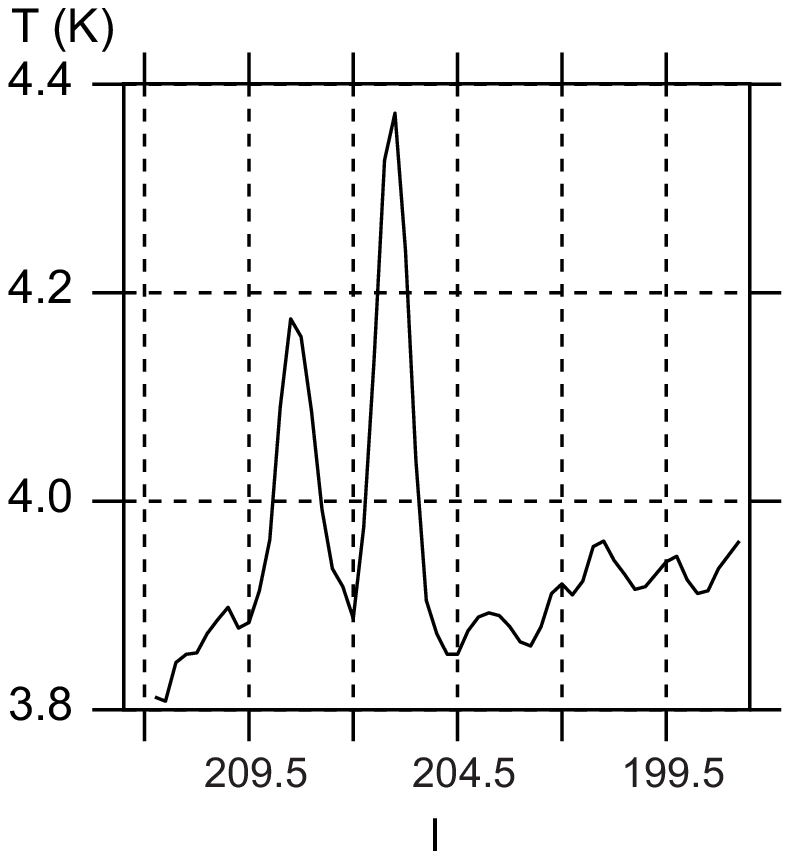}
\caption{Temperature profiles at 1420 MHz for Galactic longitude
from 212$^\circ$ to 198$^\circ$ and for the following values of
Galactic latitude: 1.$^\circ$5 (top left), 1$^\circ$ (top right),
0$^\circ$ (middle), -2$^\circ$ (bottom left) and -3$^\circ$ (bottom
right). Temperatures are given in K, and Galactic longitudes in
degrees.}
\label{fig01}
\end{figure*}

The area of Monoceros loop is very difficult to determine precisely
due to great influence of background radiation and superposed
external sources, such as Rosette Nebula (\cite{grah82};
\cite{ahar04,ahar07}; \cite{kim07}). Therefore, some authors made
very rough estimates for the shape of this loop, representing it by
fitted circles (see e.g. \cite{ahar04}; \cite{kim07}). In order to
make better estimates for the loop boundaries, we analyzed
temperature profiles (see Figure \ref{fig01}) for different values
of Galactic latitude ($b$) between -6$^\circ$ and 6$^\circ$, and for
Galactic longitude ($l$) from 212$^\circ$ to 198$^\circ$. As one can
see in Figure \ref{fig01}, there are two temperature peaks. The
lower one corresponds to the total brightness temperature of the
loop (including background radiation), while the higher one
corresponds to superposed external sources. We denoted the minimum
and maximum brightness temperature of the lo\-wer peak by
$T_\mathrm{min}$ and $T_\mathrm{max}$, respectively. These
temperature limits enable us to distinguish the loop from background
and also from external sources, and their values are given in Table
\ref{tab01}. Moreover, using the temperature profiles we found that
the Monoceros loop is located in this sky region: $l \in [210^\circ,
200^\circ]$ and $b \in [-6^\circ, 5^\circ]$.

\begin{figure*}
\centering
\includegraphics[width=0.48\textwidth]{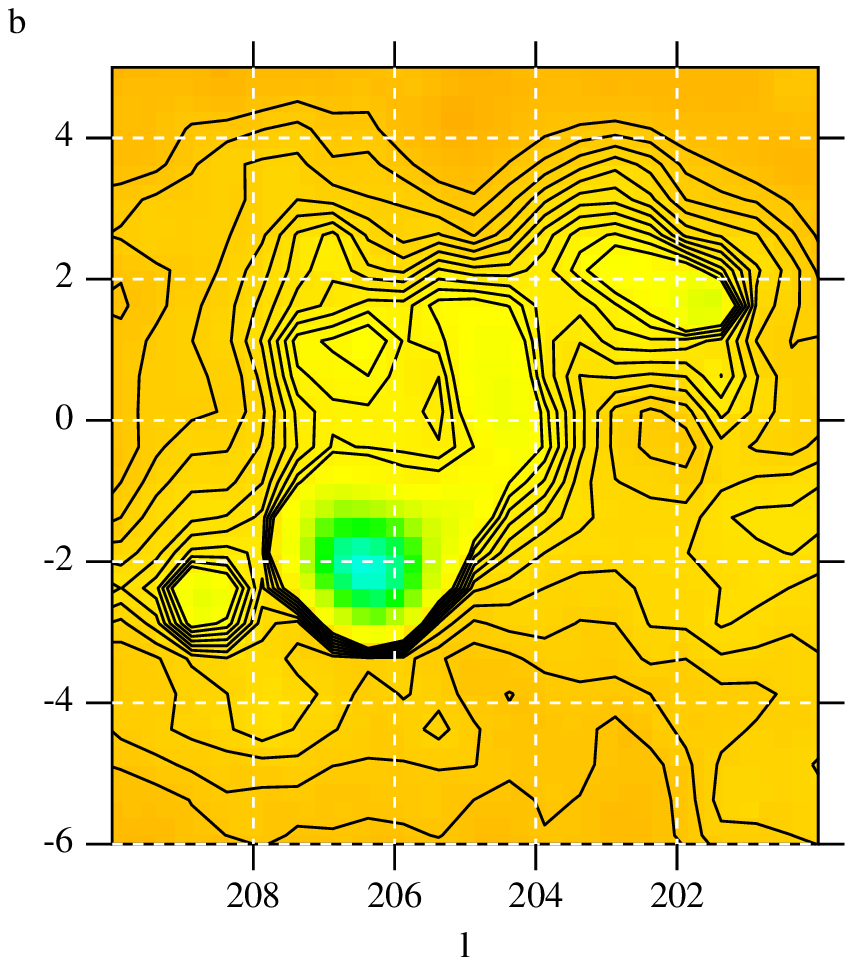}
\includegraphics[width=0.48\textwidth]{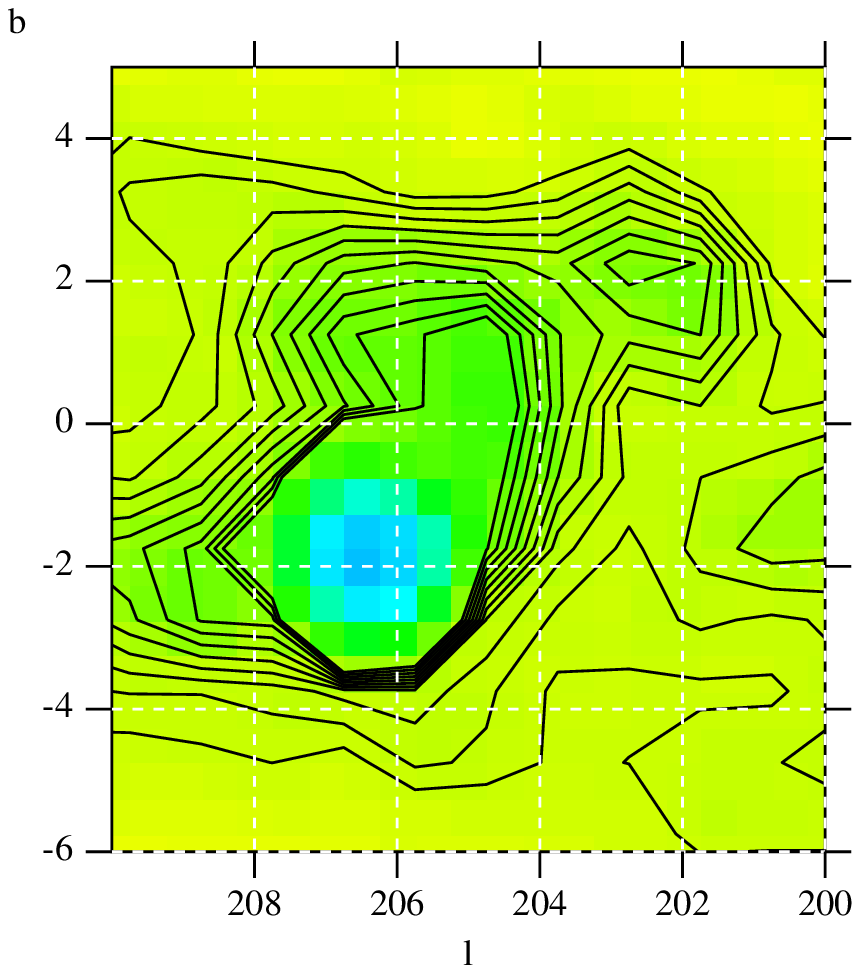}
\includegraphics[width=0.48\textwidth]{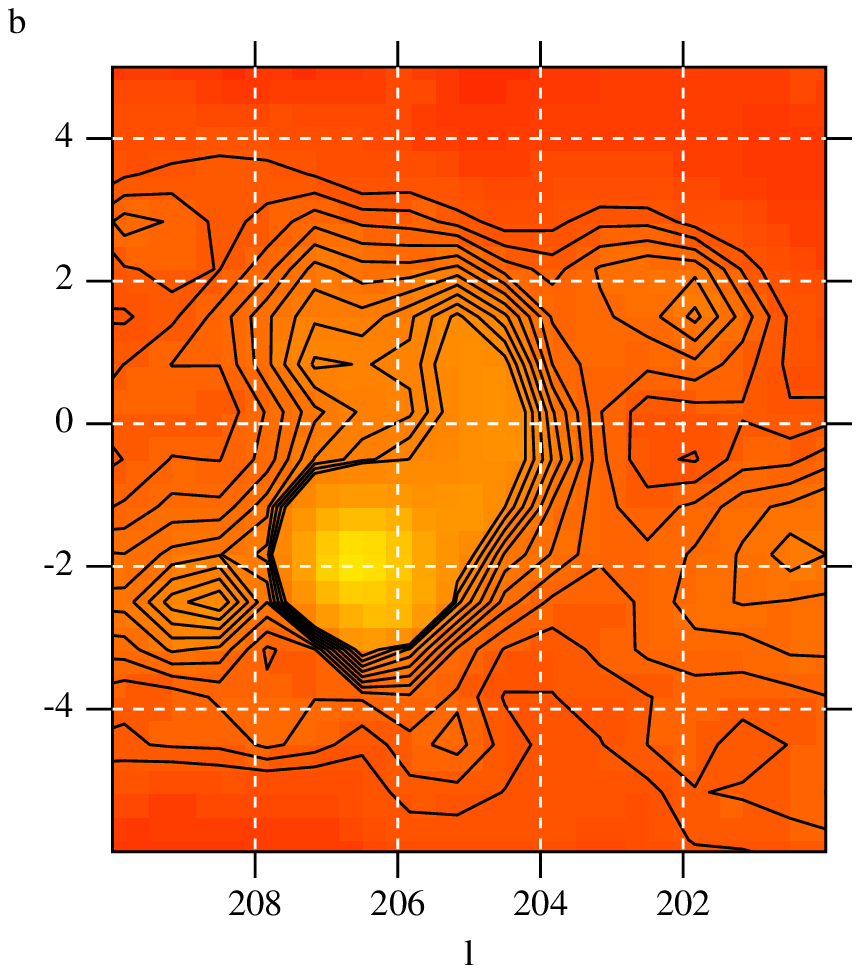}
\includegraphics[width=0.45\textwidth]{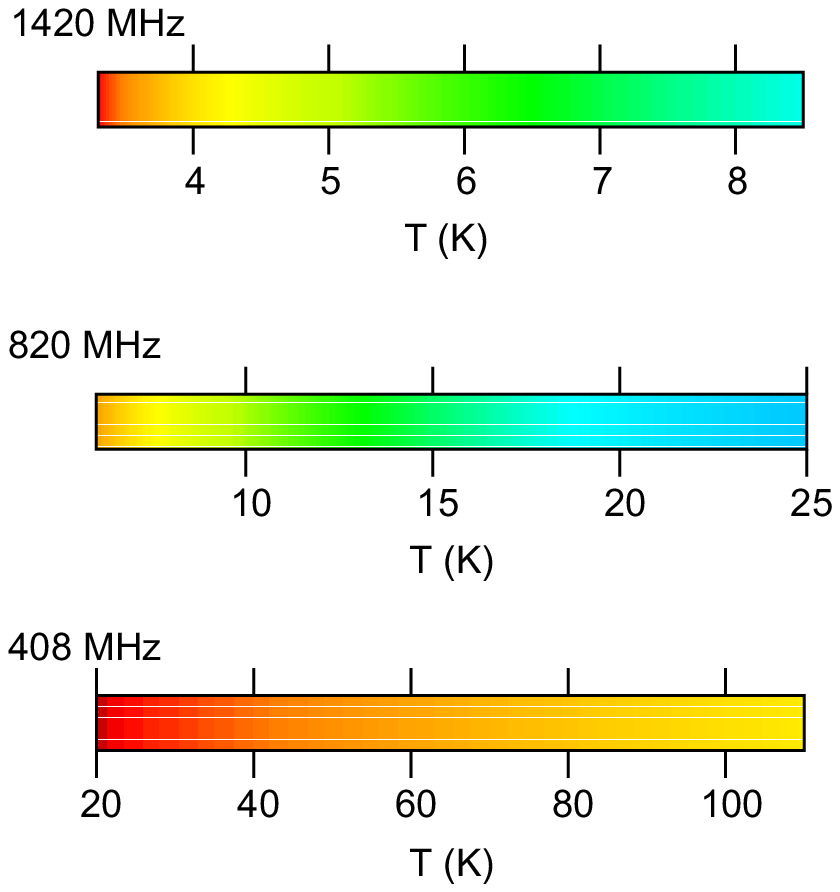}
\caption{The map of a region in Monoceros, in new galactic
coordinates (l, b). Contours of the brightness temperatures
$T_\mathrm{b}$ are plotted. \textbf{Top left:} the 1420 MHz map with
contours of $T_\mathrm{b}$ given in units of K. The contours are
plotted every 0.04 K starting from the lowest temperature of 3.8 K
up to 4.2 K. \textbf{Top right:} the 820 MHz map with $T_\mathrm{b}$
in K. The contours are plotted every 0.19 K starting from 8.8 K up
to 10.7 K. \textbf{Bottom left:} the 408 MHz map with $T_\mathrm{b}$
in K. The contours are plotted every 1.1 K starting from 36 K up to
47 K. \textbf{Bottom right:} temperature colorbars for 1420, 820 and
408 MHz, all in K.}
\label{fig02}
\end{figure*}

The area of Monoceros loop is enclosed with brightness temperature
contours (see Figure \ref{fig02}). These contour lines correspond to
the minimum and maximum brightness temperatures which define its
borders (see Table \ref{tab01}), and 9 contours in between. We have
subtracted the background radiation, as well as the superposed
radiation, in order to derive the mean brightness temperature of the
SNR alone (see \cite{borka07} and \cite{borka08}).

For deriving temperatures over the Monoceros loop, the areas used
for the individual spurs were obtained from the radio continuum maps
at three frequencies: 1420 MHz (\cite{reic86}), 820 MHz
(\cite{berk72}) and 408 MHz (\cite{hasl82}). The areas over which an
average brightness temperature is determined at each of the three
frequencies are taken to be as similar as possible within the limits
of measurement accuracy. Temperature limits $T_\mathrm{min}$ and
$T_\mathrm{max}$, given in Table \ref{tab01}, are the lower and the
upper temperature values: $T_\mathrm{min}$ is the lower temperature
limit between the background and the spur, while $T_\mathrm{max}$ is
the upper temperature limit between the spur and superposed
confusing sources. In this manner, background radiation was
considered as radiation that would exist if there were no spurs. If
the value of $T_\mathrm{min}$ ($T_\mathrm{max}$) is changed by some
small value (approximately 0.1 K for 1420 MHz and more for other
frequencies), the brightness contours become significantly
different. For evaluating brightness temperature of the background,
we used all measured values below $T_\mathrm{min}$, inside the
corresponding intervals of $l$ and $b$, and lying on the outer side
of a loop. The value of $T_\mathrm{b}$ is approximately constant
near a loop. For evaluating the brightness temperature of a loop
including the background, we used all measured values between
$T_\mathrm{min}$ and $T_\mathrm{max}$ inside the corresponding
regions of $l$ and $b$. Mean brightness temperature for loop is
found by subtracting the mean value of background brightness
temperature from the mean value of the brightness temperature over
the area of the loop. After deriving the temperature, we derived the
surface brightness by:

\begin{equation}
\Sigma = (2k \nu^2 /c^2) T_{\rm b}
\label{equ01}
\end{equation}

\noindent where $k$ is Boltzmann constant and $c$ is the speed of
light. Results are given in Table \ref{tab01}.

The surface brightnesses of SNRs must be above the sensitivity limit
of the observations and must be clearly distinguishable from the
Galactic background emission (\cite{gree91}). Therefore, the data
from the fainter parts of the loops (which are very low surface
brightness SNRs) are not taken into account because it is very
difficult to resolve them from the background. On the other hand,
this would significantly reduce brightness of entire loop and there
is a general trend that fainter SNRs tend to be larger
(\cite{gree05}). For evaluation brightness temperatures over the
spurs we had to take into account background radiation
~(\cite{webs74}). Borders enclosing the spurs are defined to
separate the spur and its background.

The measured data have different resolutions for different
frequencies (see \S 2.1), and therefore in order to obtain $T-T$
plots the data are retabulated so the higher resolution maps are
convolved to the resolution of the lowest resolution map. In that
way we convolved data at 1420 and 408 MHz to $0^\circ.5 \times
0^\circ.5$ resolution, which is the sampling rate of the 820 MHz
survey. These retabulated data are presented in Figure \ref{fig04}
for the following frequencies: 1420 MHz (top left), 820 MHz (top
right), 408 MHz (bottom left). Then, for each frequency pair we used
only the common points (with the same $(l,b)$) which belong to the
loop area at both frequencies (see Figure \ref{fig04}, bottom
right). In that way we reduced loop area to the same area for
different frequencies. The obtained $T-T$ plots for three pairs of
frequencies (between 1420-820, 1420-408 and 820-408 MHz) enabled
calculating the spectral indices. For each of the three frequency
pairs, by interchanging the dependent and independent variables we
have obtained two $\beta$ values for each pair and the mean value of
these fit results is adopted as the radio spectral index, as
suggested in \cite{uyan04}.

\section{Results and discussion}

The average brightness temperatures and surface bri\-ghtnes\-ses of
Monoceros radio loop at the three frequencies are given in Table
\ref{tab01}.

\begin{table*}
\centering
\caption{Temperatures and brightnesses of Monoceros radio
loop at 1420, 820 and 408 MHz.}
\begin{tabular}{c c c c}
\hline
Frequency & Temperature limits & Temperature & Brightness \\
(MHz) & $T_\mathrm{min}$, $T_\mathrm{max}$ (K) & (K) & (10$^{-22}$ W/(m$^2$ Hz Sr)) \\
\hline
1420 & 3.8, 4.2 & 0.18 $\pm$ 0.05 & 1.09 $\pm $ 0.30 \\
820 & 8.8, 10.7 & 0.90 $\pm$ 0.20 & 1.85 $\pm$ 0.40 \\
408 & 36, 47 & 5.2 $\pm$ 1.0 & 2.63 $\pm$ 0.50 \\
\hline
\end{tabular}
\label{tab01}
\end{table*}

Spectrum shows how temperature depends on the frequency:

\begin{equation}
\log T = \log K - \beta \log \nu
\label{equ02}
\end{equation}

\noindent where $K$ is a constant, and $\beta$ is radio spectral
index.

\begin{figure*}
\centering
\includegraphics[width=0.48\textwidth]{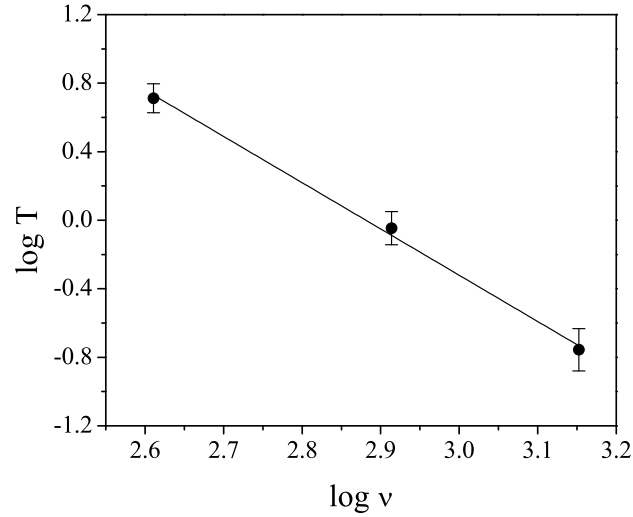}
\caption{Monoceros loop spectrum: temperature versus frequency, for
three measurements -- at 408, 820 and 1420 MHz.}
\label{fig03}
\end{figure*}

\begin{figure*}
\centering
\includegraphics[width=0.48\textwidth]{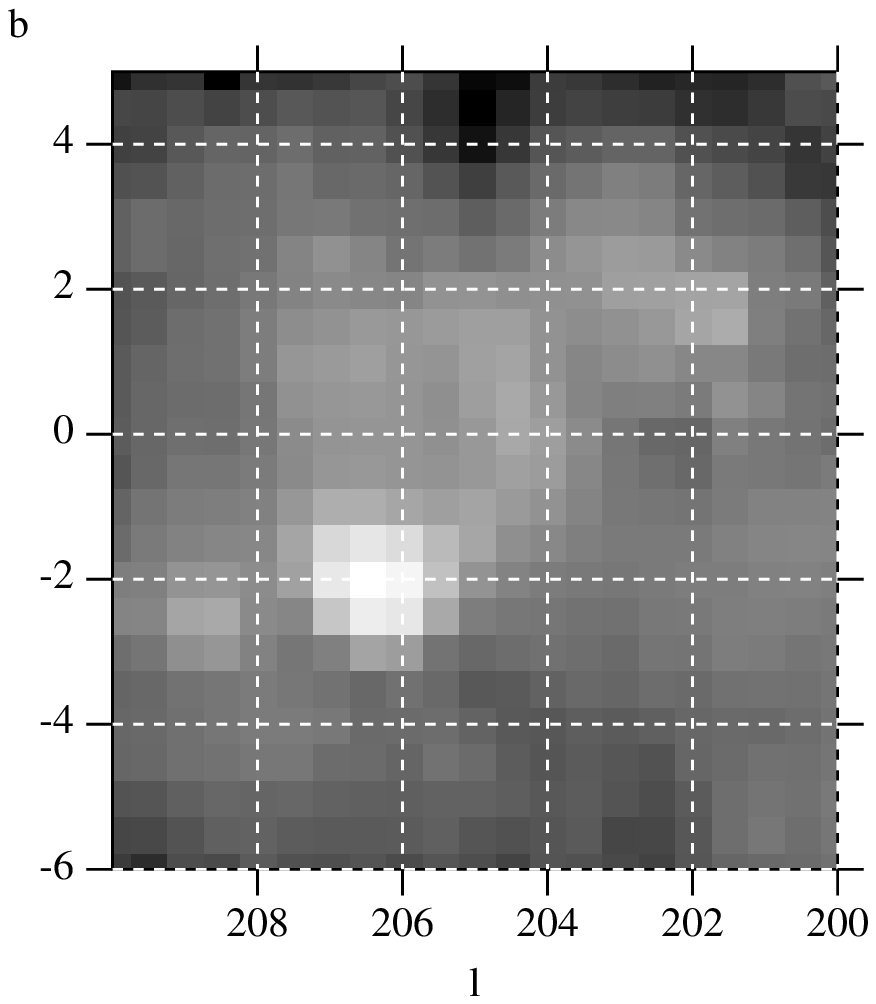}
\includegraphics[width=0.48\textwidth]{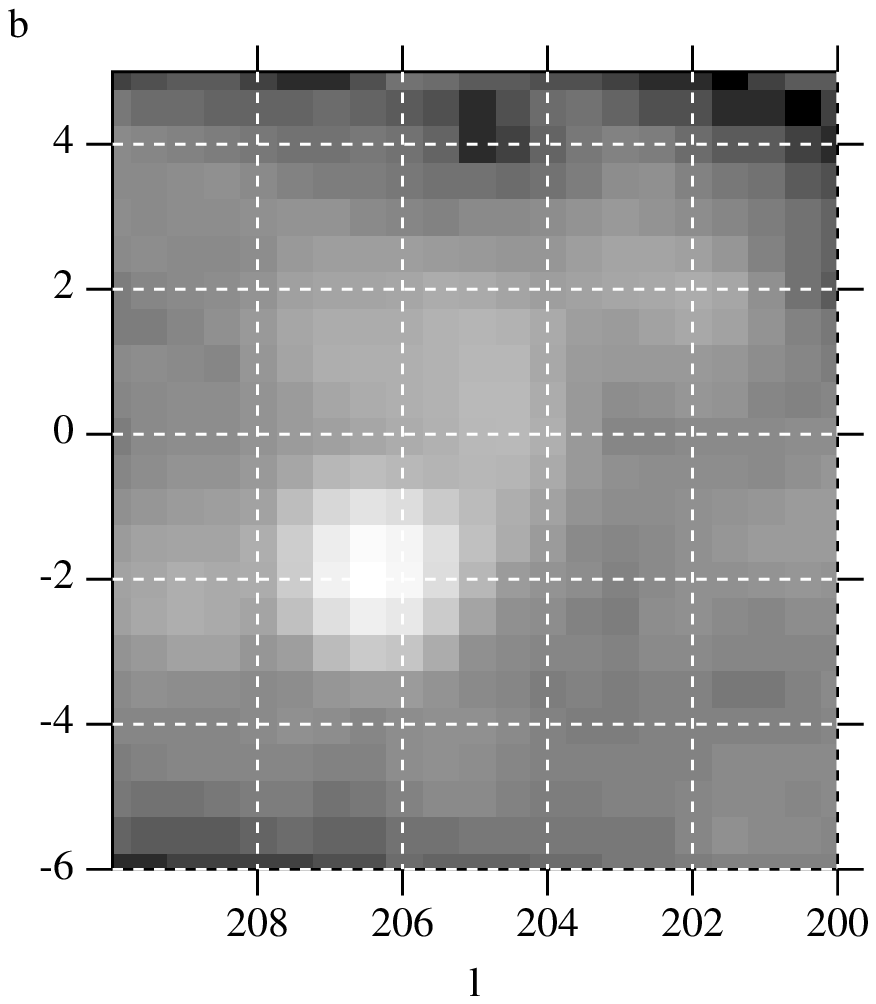}
\includegraphics[width=0.48\textwidth]{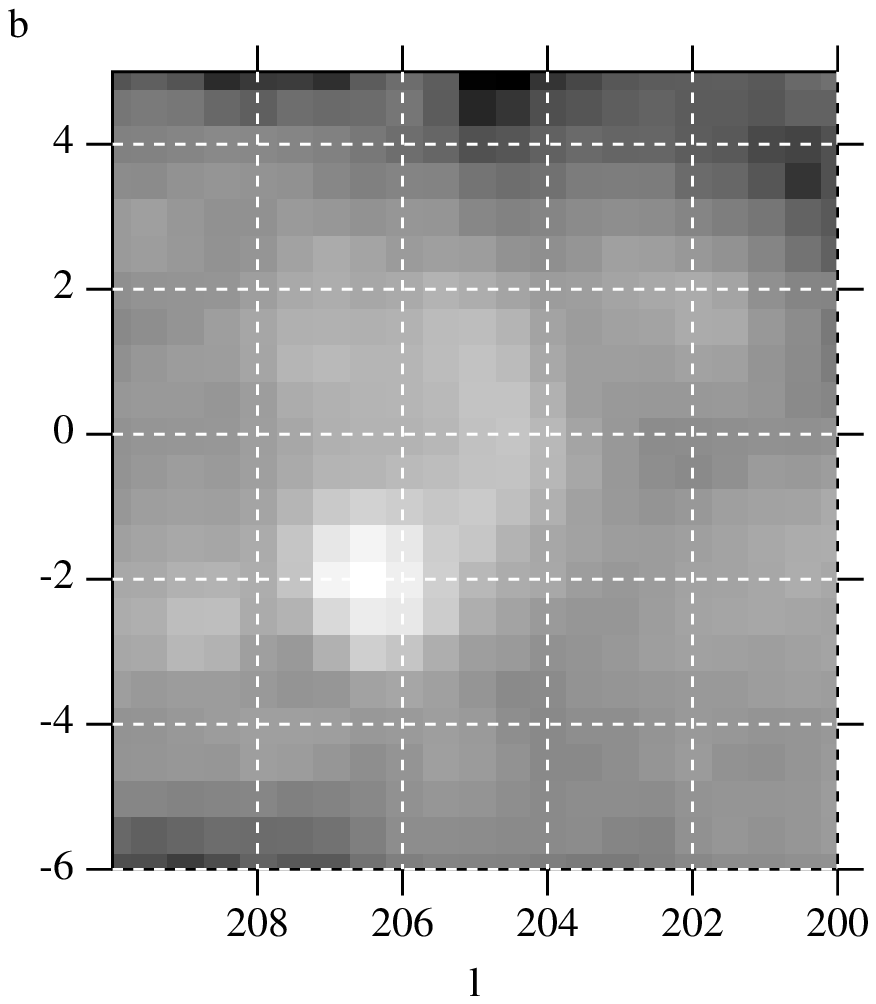}
\includegraphics[width=0.48\textwidth]{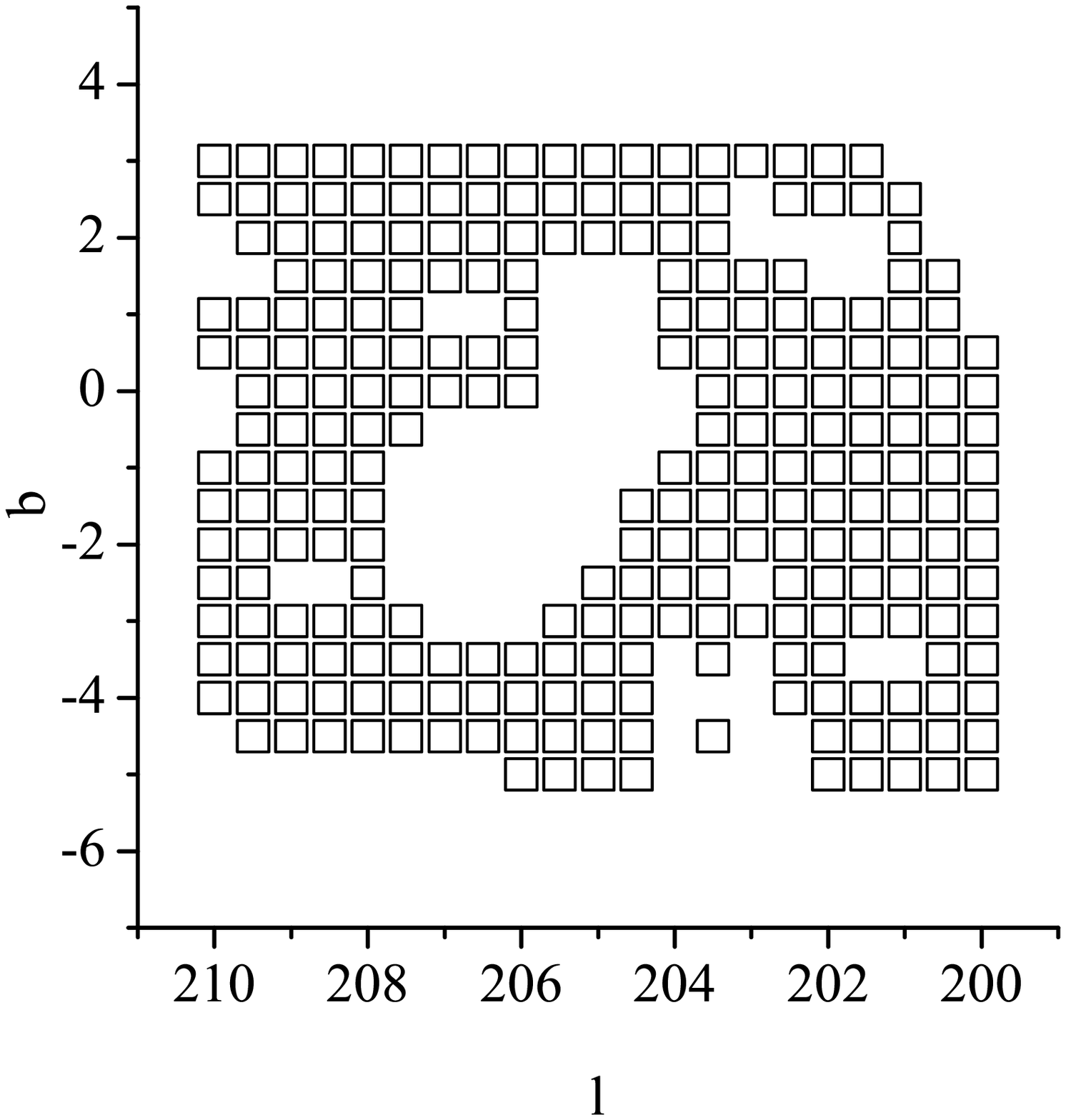}
\caption{The data retabulated to $0^\circ.5 \times 0^\circ.5$
resolution, for the following frequencies: 1420 MHz \textbf{(top
left)}, 820 MHz \textbf{(top right)} and 408 MHz \textbf{(bottom
left)}. \textbf{Bottom right:} the common pixels which belong to the
loop area at all frequencies.}
\label{fig04}
\end{figure*}

The spectrum was generated using mean temperatures at three
different frequencies. This best-fit straight line spectrum enables
calculation of spectral index as negative value of the line's
direction coefficient. The spectrum is shown in Figure \ref{fig03}.
Relative errors of the measurements $\Delta \log T = \frac{{\Delta
T}}{{T \ln 10}}$ are presented by error bars, where $\Delta T$ are
the corresponding absolute errors given in Table \ref{tab01}.
Obtained value $\beta$ = 2.70 $\pm$ 0.14 (greater than 2.2) confirm
non-thermal origin of Monoceros loop emission. The value for the
brightness temperature spectral index of the Monoceros loop is
rather steep (about 2.7). This is at the high end of the spectral
index distribution for SNRs as suggested in \cite{clar76}.

From the relation ($T-T$ plot):

\begin{equation}
T_{\nu 1} / T_{\nu 2} = (\nu_1 / \nu_2)^{-\beta}
\label{equ03}
\end{equation}

\noindent $\beta$ can be determined as

\begin{equation}
\beta = \log(a_{12}) / \log(\nu_1 / \nu_2),
\label{equ04}
\end{equation}

\noindent where $a_{12}$ is direction coefficient of line $T_{\nu
2}(T_{\nu 1})$.

\begin{figure*}
\centering
\includegraphics[width=0.48\textwidth]{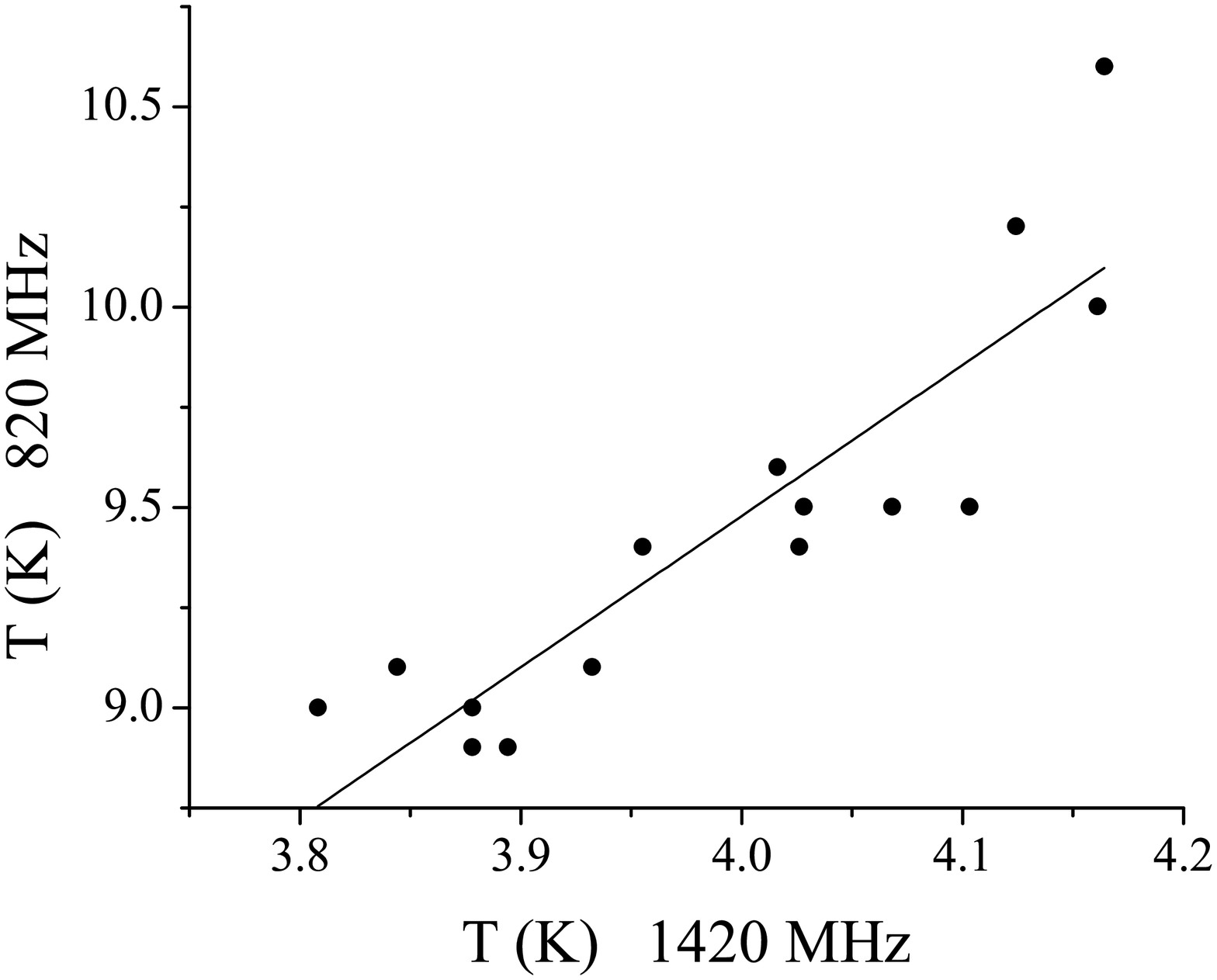}
\includegraphics[width=0.48\textwidth]{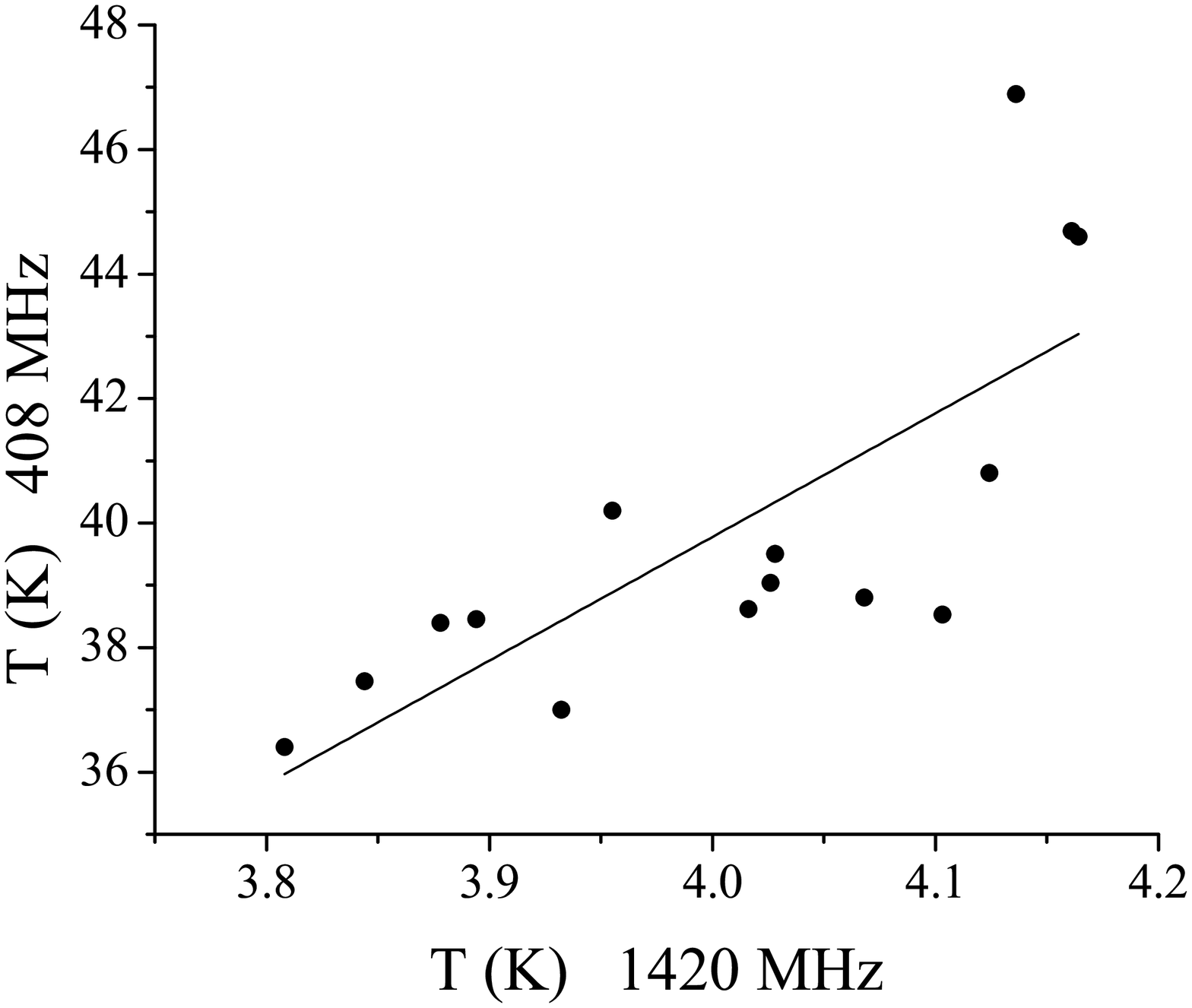}
\includegraphics[width=0.48\textwidth]{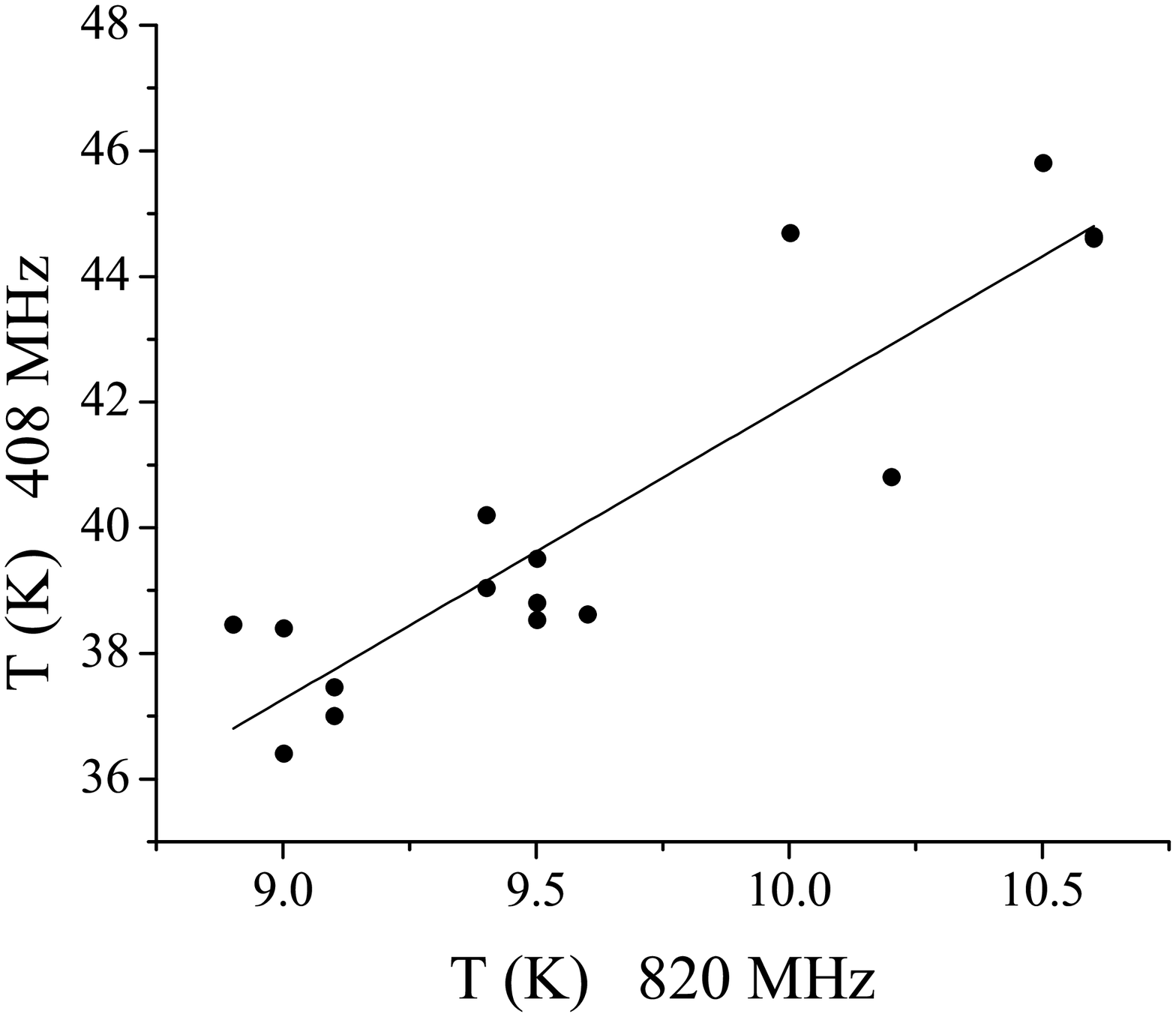}
\caption{\textbf{Top left:} T-T plot between 1420 and 820 MHz for
Monoceros loop at latitude $b$ = 1$^\circ$, in longitude range
[210$^\circ$, 200$^\circ$]. \textbf{Top right:} the same as top
left, but between 1420 and 408 MHz. \textbf{Bottom:} the same as top
left, but between 820 and 408 MHz.}
\label{fig05}
\end{figure*}

Because of the influence of superposed sources to the loop and
because it is hard to estimate precise loop's borders, we present
how spectral indices vary for different $b$. Variations of spectral
indices between the three frequency pairs obtained from T-T plots,
distributed over Galactic latitudes, are given in Tables
\ref{tab02}-\ref{tab04}. For some Galactic latitudes, there is a
large dispersion of the points in $T-T$ graphs so they are not
suitable for calculation of spectral index; we did not calculate
$\beta$ for them and it is denoted with "-". The averaged values of
radio spectral indices are given in Table \ref{tab05}. Examples of
$T-T$ plots between 1420-820, 1420-408 and 820-408 MHz for longitude
$b$ = 1$^{\circ}$ are given in Figure \ref{fig05}. The average value
of spectral index from $T-T$ is $<\beta_{TT}>$ = 2.63 $\pm$ 0.30. It
can be noticed that this value agrees well with the corresponding
value obtained from spectrum, as expected (see \cite{uyan04}). Then
we calculated mean value of spectral index between 1420, 820 and 408
MHz (regarding spectrum and $T-T$ graphs) $<\beta>$ = 2.66 $\pm$
0.20. With this $<\beta>$ we reduced $\Sigma_{\nu}$ to 1000 MHz
(Table \ref{tab06}) according to relation:

\begin{equation}
\Sigma_\mathrm{1 GHz} / \Sigma_{\nu \mathrm{GHz}} = (\nu /
1)^{(\beta-2)}.
\label{equ05}
\end{equation}

\begin{table}
\centering
\caption{Variations of spectral indices between 1420 and 820 MHz
obtained from T-T plots, distributed over Galactic latitudes.}
\begin{tabular}{c c c}
\hline
latitude ($^\circ$) & $\beta(1420-820)$ & $\beta(820-1420)$ \\
\hline
-5.0 & - & - \\
-4.5 & - & - \\
-4.0 & - & - \\
-3.5 & - & - \\
-3.0 & 1.92 $\pm$ 0.82 & 4.54 $\pm$ 0.82 \\
-2.5 & 3.86 $\pm$ 0.46 & 4.84 $\pm$ 0.46 \\
-2.0 & 3.47 $\pm$ 0.31 & 4.07 $\pm$ 0.31 \\
-1.5 & 2.96 $\pm$ 0.73 & 5.02 $\pm$ 0.73 \\
-1.0 & 1.96 $\pm$ 0.55 & 3.31 $\pm$ 0.55 \\
-0.5 & 2.61 $\pm$ 0.39 & 3.36 $\pm$ 0.39 \\
0 & 2.39 $\pm$ 0.22 & 2.73 $\pm$ 0.22 \\
0.5 & 2.49 $\pm$ 0.28 & 3.09 $\pm$ 0.28 \\
1.0 & 2.42 $\pm$ 0.26 & 2.85 $\pm$ 0.26 \\
1.5 & 2.74 $\pm$ 0.27 & 3.13 $\pm$ 0.27 \\
2.0 & 2.62 $\pm$ 0.26 & 3.02 $\pm$ 0.26 \\
2.5 & 1.59 $\pm$ 0.41 & 2.69 $\pm$ 0.41 \\
3.0 & 1.26 $\pm$ 0.43 & 2.41 $\pm$ 0.43 \\
\hline
\end{tabular}
\label{tab02}
\end{table}

\begin{table}
\centering
\caption{Variations of spectral indices between 1420 and
408 MHz obtained from T-T plots, distributed over Galactic
latitudes.}
\begin{tabular}{c c c}
\hline
latitude ($^\circ$) & $\beta(1420-408)$ & $\beta(408-1420)$ \\
\hline
-5.0 & 1.92 $\pm$ 0.24 & 2.32 $\pm$ 0.24 \\
-4.5 & 2.14 $\pm$ 0.33 & 3.15 $\pm$ 0.33 \\
-4.0 & 2.35 $\pm$ 0.12 & 2.62 $\pm$ 0.12 \\
-3.5 & - & - \\
-3.0 & 2.48 $\pm$ 0.19 & 3.03 $\pm$ 0.19 \\
-2.5 & 2.91 $\pm$ 0.13 & 3.14 $\pm$ 0.13 \\
-2.0 & 2.58 $\pm$ 0.11 & 2.78 $\pm$ 0.11 \\
-1.5 & 2.55 $\pm$ 0.14 & 2.82 $\pm$ 0.14 \\
-1.0 & 2.55 $\pm$ 0.08 & 2.65 $\pm$ 0.08 \\
-0.5 & 2.69 $\pm$ 0.07 & 2.76 $\pm$ 0.07 \\
0 & 2.58 $\pm$ 0.10 & 2.75 $\pm$ 0.10 \\
0.5 & 2.39 $\pm$ 0.16 & 2.78 $\pm$ 0.16 \\
1.0 & 2.40 $\pm$ 0.19 & 2.82 $\pm$ 0.19 \\
1.5 & 2.16 $\pm$ 0.22 & 2.65 $\pm$ 0.22 \\
2.0 & 2.19 $\pm$ 0.25 & 2.80 $\pm$ 0.25 \\
2.5 & - & 3.23 $\pm$ 0.89 \\
3.0 & - & - \\
\hline
\end{tabular}
\label{tab03}
\end{table}

\begin{table}
\centering
\caption{Variations of spectral indices between 820 and
408 MHz obtained from T-T plots, distributed over Galactic
latitudes.}
\begin{tabular}{c c c}
\hline
latitude ($^\circ$) & $\beta(820-408)$ & $\beta(408-820)$ \\
\hline
-5.0 & 2.38 $\pm$ 0.51 & 3.29 $\pm$ 0.51 \\
-4.5 & 1.39 $\pm$ 0.89 & 4.20 $\pm$ 0.89 \\
-4.0 & 1.71 $\pm$ 0.49 & 3.22 $\pm$ 0.49 \\
-3.5 & 1.12 $\pm$ 0.56 & 2.74 $\pm$ 0.56 \\
-3.0 & 1.87 $\pm$ 0.33 & 2.74 $\pm$ 0.33 \\
-2.5 & 1.57 $\pm$ 0.28 & 2.07 $\pm$ 0.28 \\
-2.0 & 1.47 $\pm$ 0.25 & 1.95 $\pm$ 0.25 \\
-1.5 & 1.24 $\pm$ 0.35 & 2.07 $\pm$ 0.35 \\
-1.0 & 2.23 $\pm$ 0.31 & 2.87 $\pm$ 0.31 \\
-0.5 & 2.44 $\pm$ 0.17 & 2.67 $\pm$ 0.17 \\
0 & 2.49 $\pm$ 0.10 & 2.58 $\pm$ 0.10 \\
0.5 & 2.23 $\pm$ 0.12 & 2.36 $\pm$ 0.12 \\
1.0 & 2.22 $\pm$ 0.18 & 2.50 $\pm$ 0.18 \\
1.5 & 1.91 $\pm$ 0.25 & 2.36 $\pm$ 0.25 \\
2.0 & 1.96 $\pm$ 0.18 & 2.29 $\pm$ 0.18 \\
2.5 & - & 3.21 $\pm$ 0.72 \\
3.0 & - & - \\
\hline
\end{tabular}
\label{tab04}
\end{table}

\begin{table}
\centering
\caption{The average radio spectral indices between 1420,
820 and 408 MHz obtained from T-T plots.}
\begin{tabular}{c c}
\hline
Frequency pair (MHz) & $\beta_{TT}$ \\
\hline
1420 - 820 & 2.98 $\pm$ 0.40 \\
1420 - 408 & 2.62 $\pm$ 0.20 \\
820 - 408 & 2.29 $\pm$ 0.34 \\
\hline
\end{tabular}
\label{tab05}
\end{table}

In \cite{grah82}, the value $\beta$ = 2.47 $\pm$ 0.06 of radio
spectral index of this loop was derived from the radio continuum
spectrum between two frequencies: 111 and 2700 MHz, and we
calculated spectrum at three frequencies and also derived $T-T$
graphs which represent different method.

\begin{table}
\centering
\caption{Brightnesses of Monoceros radio loop reduced to
1000 MHz using $\beta$ = 2.66 $\pm$ 0.20.}
\begin{tabular}{c c}
\hline
Frequency & Brightness at 1000 MHz\\
(MHz) & (10$^{-22}$ W/(m$^2$ Hz Sr)) \\
\hline
1420 & 1.37 $\pm $ 0.50 \\
820 & 1.63 $\pm $ 0.30 \\
408 & 1.45 $\pm $ 0.03 \\
\hline
\end{tabular}
\label{tab06}
\end{table}

Distances to the SNRs can be inferred from positional coincidences
with HI, HII regions and molecular clouds, OB associations, or
pulsars or from measuring optical velocities and proper motions.
Where there is no direct distance determination, estimates can be
made for shell remnants by utilizing the radio surface
brightness-to-diameter relationship ($\Sigma-D$) (\cite{case98} and
references therein). The mean surface brightness at a specific radio
frequency, $\Sigma_{\nu}$ is a distance-independent parameter and,
to a first approximation, is an intrinsic property of the SNR
(\cite{shkl60}).

Using a set of 172 SNRs as calibrators, for which reliable distance
values were determined, \cite{uros02} has constructed the following
relation between the surface brig\-htness ($\Sigma$) and diameter
($D$):

\begin{equation}
\Sigma_\mathrm{1GHz} = 2.76 \times 10^{-16} D^{-3.02}
\label{equ06}
\end{equation}

\noindent This relation, combining the Galactic and ~ extragalactic
SNRs (master relation), also includes four main Galactic radio loops
(Loops I -- IV).

It is shown (\cite{odeg86}) that there is diffuse emission with the
filaments located in the western half of Monoceros loop (overlapping
with Rosette Nebula) and in the eastern part (dust cloud L1631).
There is also a peak in the 2700 MHz emission in the eastern edge
for which \cite{grah82} suggested to be a HII region. In
\cite{ahar07} it is said that the complex Monoceros loop/Rosette
Nebula region contains several potential sources of very-high-energy
$\gamma$ ray emission and that the interaction of the SNR with a
compact molecular cloud is possible. Because of possible influence
with molecular clouds and HII region, we also take $\Sigma - D$
relation from \cite{arbu04} adjusted for molecular clouds:

\begin{equation}
\Sigma_\mathrm{1GHz} = 1.1 \times 10^{-15} D^{-3.5}
\label{equ07}
\end{equation}

\noindent They have taken 14 shell-type Galactic SNRs associated
with large molecular clouds.

Using the relations derived by \cite{uros02} and \cite{arbu04}, we
computed diameter of the loop and then we computed distance as:

\begin{equation}
r = D / (2 \sin \Theta),
\label{equ08}
\end{equation}

\noindent with angular radius ($\Theta$) taken from \cite{uros98}.
The results are given in Table \ref{tab07}.

It can be noticed that there is the influence of molecular cloud on
Monoceros loop (\cite{grah82,ahar07}). If that influence constantly
existed during the loop's evolution, then for distance calculation
it is more suitable $\Sigma - D$ relation by \cite{arbu04}. If not,
then it is more suitable to use relation by \cite{uros02}.

\begin{table}
\centering
\caption{Diameters $D$ (pc) and distances $r$ (pc) of
Monoceros radio loop derived from the $\Sigma-D$ relation given by
\cite{uros02} and by \cite{arbu04}.}
\begin{tabular}{c c c}
\hline
Relation & $D$ (pc) & $r$ (pc) \\
\hline
(\ref{equ06}) & 119 $\pm$ 18 & 1630 $\pm$ 250 \\
(\ref{equ07}) & 92 $\pm$ 14 & 1250 $\pm$ 190 \\
\hline
\end{tabular}
\label{tab07}
\end{table}

For the distance to this loop, mean optical velocity suggests 0.8
kpc, and low frequency radio absorption suggests 1.6 kpc
(\cite{grah82,odeg86,gree06}). With our calculated radio spectral
index $\beta$ = 2.66 $\pm$ 0.20, we calculated distance to this loop
using our derived brightnesses. The results are given in Table
\ref{tab07}.

\section{Conclusions}

In this paper we calculated the brightness temperatures and surface
brightnesses of the Monoceros radio loop at 1420, 820 and 408 MHz.
Linear spectrum is estimated for mean temperatures versus frequency
between 1420, 820 and 408 MHz. It is the first time that the
brightness temperatures of Monoceros loop are calculated at 820 and
408 MHz frequencies from the observational data. We sampled much
more points (more than 1~000) at 1420 MHz than in previous papers
(95 points) (\cite{uros98,jmt96}). Also, the brightness temperature
is now derived using the different method. The temperature of this
radio loop at 1420 MHz is in good agreement with the result obtained
in \cite{uros98}.

We present the radio continuum spectrum of the Monoceros loop using
average brightness temperatures at three different frequencies. As
it can be seen from Figure \ref{fig03}, given linear
fit provides reliable spectral index. Also, we present the
$T-T$ plots which enables also calculation of spectral index.

The effective sensitivity of the brightness temperatures are: 50 mK
for 1420 MHz, 0.2 K for 820 MHz, and about 1.0 K for
408 MHz. The most precise measurements (the least relative errors)
are in case of 1420 MHz, so positions of the brightness temperature
contours of the loop are the most realistic for this frequency.
The brightnesses of Monoceros loop at 1420 and 820 and 408 MHz are in
good agreement when reduced to 1 GHz.

With our derived brightnesses, we calculated new diameters and
distances to this loop at the three frequencies: 1420, 820 and 408
MHz and then estimated some average distance. We used empirical
$\Sigma-D$ relations for supernova remnants by \cite{uros02} and
\cite{arbu04}. The estimated distance of the Monoceros radio loop is
in good agreement with the earlier results (\cite{guse04};
\cite{gree06}: 1.6 kpc).

The spectral index analysis confirms non-thermal origin of Monoceros
radio loop.

\acknowledgements

This research is part of the projects: "Gaseous and stellar
component of galaxies: interaction and evolution" (No. 146012) and
"Physics and chemistry with ion beams" (No. 451-01-00049) supported
by the Ministry of Science of Republic of Serbia.

\end{document}